\documentclass{PoS}
\usepackage{amsmath}
\usepackage{subfig}
\usepackage{multicol}
\usepackage{graphicx}
\title{$K$ and $D$ oscillations in the Standard Model and its extensions from $N_f=2+1+1$ Twisted Mass LQCD}

\ShortTitle{$K$ and $D$ oscillations from $N_f=2+1+1$ Twisted Mass LQCD}

\vspace{-0.23cm}
\author{\speaker{N.~Carrasco}, V.~Gim\'enez\\
        Dep. de F\'isica Te\`orica and IFIC, Universitat de Val\`encia-CSIC\\
\vspace{-0.23cm}
}
\vspace{-0.23cm}
\author{P.~Dimopoulos \\
        Centro Fermi - Museo Storico della Fisica e Centro Studi e Ricerche Enrico Fermi\\
        Dip. di Fisica, Universit\`a di Roma "Tor Vergata" and INFN "Tor Vergata"\\
\vspace{-0.23cm}
}
\vspace{-0.23cm}
\author{R.~Frezzotti,  G.~C.~Rossi \\
        Dip. di Fisica, Universit\`a di Roma "Tor Vergata" and INFN "Tor Vergata"\\
\vspace{-0.23cm}
}
\vspace{-0.23cm}
\author{V.~Lubicz, C.~Tarantino\\
        Dip. di Fisica, Universit\`a Roma Tre and INFN Roma Tre \\
\vspace{-0.23cm}
}
\vspace{-0.23cm}
\author{F.~Sanfilippo\\
       Laboratoire de Physique The\'orique, Universit\'e Paris Sud \\
\vspace{-0.23cm}
}     
\vspace{-0.23cm}
\author{S.~Simula\\
        INFN Roma Tre \\
\vspace{-0.23cm}
}

\abstract{

We present the first $N_f=2+1+1$ results for the matrix elements of the operators describing neutral $K$ and $D$ mixing in the Standard Model and its extensions. The combination of maximally twisted sea quarks and Osterwalder-Seiler valence quarks ensures $\mathcal{O}(a)$-improvement and continuum like renormalization pattern. We have used the $N_f=2+1+1$ dynamical quark gauge configurations generated by ETMC. Simulations include three  lattice spacings in the interval $[0.06:0.09]$ fm and pseudoscalar  meson masses in the range $[230:500]$ MeV. Our results are extrapolated to the continuum limit and to the physical quark masses. 
The calculation of the renormalization constants has been performed non-perturbatively in the RI-MOM scheme. 
}

\FullConference{31st International Symposium on Lattice Field Theory - LATTICE 2013\\
		July 29 - August 3, 2013\\
		Mainz, Germany}

\begin{document}

\section{Introduction}
\vspace*{-0.3cm}
We provide the first $N_f=2+1+1$ accurate lattice determination of the $\Delta S=2$ and $\Delta C=2$ bag parameters relevant for physics in the SM and beyond. In \cite{Bae:2013tca,Boyle:2012qb} recent $N_f=2+1$ results are presented. Other results reported in this conference can be found in \cite{BiPOS}.

$\Delta F=2$ processes provide some of the most stringent constraints on New Physics generalizations of the Standard Model. The most general $\Delta F=2$ effective Hamiltonian of dimension-six operators contributing to $P^0-\overline{P}^0$ meson mixing is 
\vspace*{-0.3cm}
\begin{equation}
H_{\textrm{eff}}^{\Delta F=2}={\displaystyle \sum_{i=1}^{5}C_{i}(\mu)Q_{i}},
\label{eq:Heff}
\end{equation}
\vspace*{-0.1cm}
where $C_i$ are the Wilson coefficients which encode the short distance contributions and $\mu$ is the renormalization scale. The operators $Q_i$ involving light ($\ell$) and strange or charm ($h$) quarks read, in the so-called SUSY basis,
\vspace*{-0.3cm}
\begin{equation}
\begin{array}{lll}
Q_{1}=\left[\bar{h}^{a}\gamma_{\mu}(1-\gamma_{5})\ell^{a}\right]\left[\bar{h}^{b}\gamma_{\mu}(1-\gamma_{5})\ell^{b}\right]\\
Q_{2}=\left[\bar{h}^{a}(1-\gamma_{5})\ell^{a}\right]\left[\bar{h}^{b}(1-\gamma_{5})\ell^{b}\right] &  & Q_{3}=\left[\bar{h}^{a}(1-\gamma_{5})\ell^{b}\right]\left[\bar{h}^{b}(1-\gamma_{5})\ell^{a}\right]\\
Q_{4}=\left[\bar{h}^{a}(1-\gamma_{5})\ell^{a}\right]\left[\bar{h}^{b}(1+\gamma_{5})\ell^{b}\right] &  & Q_{5}=\left[\bar{h}^{a}(1-\gamma_{5})\ell^{b}\right]\left[\bar{h}^{b}(1+\gamma_{5})\ell^{a}\right].
\label{eq:Qi}
\end{array} 
\end{equation}
\vspace*{-0.1cm}
The long-distance contributions are described by the matrix elements of the renormalized four-fermion operators. The renormalized bag parameters, $B_i$ $(i=1,...,5)$, provide the value of four-fermion matrix elements in units of the deviation from their vacuum insertion approximation. They are defined as
\begin{equation}
\begin{array}{l}
\langle\overline{P}^{0}|Q_{1}(\mu)|P^{0}\rangle=C_{1} B_{1} (\mu)m_{P}^{2}f_{P}^{2} ,\\
\langle\overline{P}^{0}|Q_{i}(\mu)|P^{0}\rangle=C_{i} B_{i}(\mu) m_{P}^{2}f_{P}^{2} \dfrac{m_{P}^{2}}{\left(m_{h}(\mu)+m_{\ell}(\mu)\right)^{2}} ,
\end{array}
\end{equation}
\vspace*{-0.1cm}
where $C_i={8/3,-5/3,1/3,2,2/3}$, $i=1,..,5$. $|P^0\rangle$ is the pseudoscalar, $K$ or $D$, state, $m_{P}$ and $f_{P}$ are the pseudoscalar mass and decay constant and $m_{h}$ and $m_{\ell}$ are the renormalized quark masses. 

\section{Lattice setup}
\vspace*{-0.3cm}
Simulations have been performed at three values of the lattice spacing using the $N_f=2+1+1$ dynamical quark configurations produced by  ETMC \cite{Baron:2011sf}. In the gauge sector, the Iwasaki action has been used while the dynamical sea quarks have been regularized employing the Twisted Mass action  \cite{Frezzotti:2000nk}  at maximal twist which ensures $\mathcal{O}(a)$ improvement  \cite{Frezzotti:2003ni,Frezzotti:2003xj}. The fermionic action for the light doublet  reads
\vspace*{-0.3cm}
\begin{equation}
S_{\ell}=\sum_{x}\bar{\psi}_{\ell}(x)\left\{ \dfrac{1}{2}\gamma_{\mu}\left(\nabla_{\mu}+\nabla_{\mu}^{*}\right)-i\gamma_{5}\tau^{3}\left[M_{\textrm{cr}}-\dfrac{a}{2}\sum_{\mu}\nabla_{\mu}^{*}\nabla_{\mu}\right]+\mu_{\textrm{sea}}\right\} \psi_{\ell}(x)
\end{equation}
\vspace*{-0.1cm}
\vspace*{-0.1cm}
where we follow the notation of \cite{Baron:2010bv}. In the heavy sector the sea quark action is
\vspace*{-0.3cm}
\begin{equation}
S_{h}=\sum_{x} \bar{\psi}_{h}(x)\left\{ \dfrac{1}{2}\gamma_{\mu}\left(\nabla_{\mu}+\nabla_{\mu}^{*}\right)- i\gamma_{5}\tau^{1}\left[m_{0}-\dfrac{a}{2}\sum_{\mu}\nabla_{\mu}^{*}\nabla_{\mu}\right]+{\mu}_{\sigma}+{\mu}_{\delta}{\tau}_{3} \right\}  {\psi}_{h}(x) .
\end{equation}
\vspace*{-0.1cm}
Continuum-like renormalization pattern for the four-fermion operators and $\mathcal{O}(a)$-improvement are achieved using a mixed action setup. We introduce Osterwalder-Seiler \cite{Osterwalder:1977pc} valence quarks allowing for a replica of the heavy $(h,h')$ and the light $(\ell,\ell')$ quarks \cite{Frezzotti:2004wz}. The valence quark action reads
\vspace{-0.35cm}
\begin{equation}
S^{\textrm{OS}}=\sum_{x}\sum_{f={\ell},{\ell}',h,h'}\bar{q}_{f}\left\{ \dfrac{1}{2}\gamma_{\mu}\left(\nabla_{\mu}+\nabla_{\mu}^{*}\right)-i\gamma_{5}r_{f}\left[M_{\textrm{cr}}-\dfrac{a}{2}\sum_{\mu}\nabla_{\mu}^{*}\nabla_{\mu}\right]+\mu_{f}\right\} q_{f}(x), 
\end{equation}
\vspace*{-0.1cm}
where the Wilson parameters are conveniently chosen such that $r_h=r_{\ell}=r_{h'}=-r_{\ell'}$ \cite{Frezzotti:2004wz}.

In table \ref{tab:simulation-detail} we give the details of the simulation and the values of the sea and the valence quark masses at each lattice spacing. The smallest sea quark mass corresponds to a pion of about 230 MeV and is attained at $\beta=2.10$. We simulate three heavy valence quark masses $\mu_{``s"}$ around the physical strange one  and  three $\mu_{``c"}$ around the physical charm mass to allow for a smooth interpolation to the physical strange and charm quark masses. For the extrapolation/interpolation to the physical quark masses we use the preliminary ETMC values \cite{LamiPOS}.

For the inversions in the valence sector we used the stochastic method with propagator sources located at random timeslices \cite{Foster:1998vw, McNeile:2006bz}. Gaussian smeared quark fields \cite{Gusken:1989qx} are implemented in the case of $D$ mesons to improve the determination of the ground state contribution with respect to the case of simple local interpolating fields. The value of the smearing parameters are $k_{G}=4$ and $N_{G}=30$. In addition, we apply APE-smearing to the gauge links in the interpolating fields \cite{Albanese:1987ds} with parameters $\alpha_{APE}=0.5$ and $N_{APE}=20$.

\vspace*{-0.3cm}
\begin{center}
\begin{table}[!h]
\begin{centering}
\scalebox{0.9}{

\begin{tabular}{cclll}
\hline 
{\small $\beta$} & {\small $L^{3}\times T$} & {\small $a\mu_{\ell}=a\mu_{\textrm{sea}}$} & {\small $a\mu_{``s"}$ } & {\small $a\mu_{``c"}$ }\tabularnewline
\hline 
{\small 1.90 ($a^{-1}\sim2.19$ GeV )} & {\small $24^{3}\times48$} &  {\small 0.0040 } & {\small 0.0145 0.0185 0.0225} &{\small 0.21256 0.25 0.29404}\tabularnewline
{\small $\mu_{\sigma}=0.15$ $\mu_{\delta}=0.19$} &    & {\small 0.0060} & &  \tabularnewline
 &  &   {\small 0.0080 } & &\tabularnewline
 &  &   {\small 0.0100} &  &\tabularnewline
\cline{2-5}
 & {\small $32^{3}\times64$} & {\small 0.0030} & {\small 0.0145 0.0185 0.0225}&{\small 0.21256 0.25 0.29404}\tabularnewline
 &    & {\small 0.0040 }  & & \tabularnewline
 &    & {\small 0.0050}  & &\tabularnewline
\hline 
{\small 1.95 ($a^{-1}\sim2.50$ GeV)} & {\small $24^{3}\times48$} & {\small 0.0085} & {\small 0.0141 0.0180 0.0219} &{\small 0.18705 0.22 0.25875 } \tabularnewline
 \cline{2-5}
{\small $\mu_{\sigma}=0.135$ $\mu_{\delta}=0.17$} & {\small $32^{3}\times64$}  & {\small 0.0025 } &  {\small 0.0141 0.0180 0.0219}&{\small 0.18705 0.22 0.25875 }\tabularnewline
 &    & {\small 0.0035 }& &\tabularnewline
 &    & {\small 0.0055} & &\tabularnewline
 &    & {\small 0.0075} & &\tabularnewline
\hline 
{\small 2.10 ($a^{-1}\sim3.23$ GeV)} & {\small $48^{3}\times96$} & {\small 0.0015 } &{\small 0.0118 0.0151 0.0184}&{\small 0.14454 0.17 0.19995}\tabularnewline
{\small $\mu_{\sigma}=0.12$ $\mu_{\delta}=0.1385$}   &  & {\small 0.0020 } & & \tabularnewline
 &    & {\small 0.0030}  & & \tabularnewline
\hline 
\end{tabular}
}
\par\end{centering}{\small \par}
\vspace*{-0.3cm}
\caption{\label{tab:simulation-detail}Simulation details} 
\vspace*{-0.3cm}
\end{table}
\vspace*{-0.3cm}
\end{center}
The computation of the renormalization constants (RCs) for the relevant two- and four-fermion operators has been performed adopting the RI'-MOM scheme \cite{Martinelli:1994ty}. These RCs are computed by extrapolating to the chiral limit the RCs estimators measured at several quark mass values. 
For the computation of the RCs, ETMC has generated dedicated runs with $N_f=4$ degenerate sea quarks.
In these $N_f=4$ simulations working at maximal twist would imply a considerable fine tuning effort to get $a m_{PCAC} \simeq 0$. 
Instead, working out of maximal twist the stability of the simulations increases. $\mathcal{O}(a)$ improvement of the RC estimators is achieved by averaging simulations 
with an equal value of the polar mass $M^{sea}$
%$\hat{M}^{sea}=Z_P^{-1} \sqrt{(Z_A m_{PCAC}^{sea})^2+(\mu^{sea})^2}$ 
but opposite value of $m_{PCAC}^{sea}$ and $\theta^{sea}$, where $\tan \theta^{sea}\, =\,  {Z_A\, m_{PCAC}^{sea}}/{\mu^{sea}}$. For details see  \cite{ETM:2011aa}.

Thanks to the OS-tm mixed action setup, the renormalized values of the bag parameters are given by the formulae \cite{Frezzotti:2004wz,Bertone:2012cu,Constantinou:2010qv}
\vspace{-0.2cm}
\begin{equation}
\begin{array}{ccc}
B_{1}=\dfrac{Z_{11}}{Z_{A}Z_{V}}B_{1}^{(b)}, & \,\,\,\,\,\,\,\,\,\,\,\,\,\,\,\,\,\,\,\,\,\, & B_{i}=\dfrac{Z_{ij}}{Z_{P}Z_{s}}B_{j}^{(b)}\,\,\,\,\,\,\,\,\,\,\,\,\, i,j=2,..,5\end{array} .
\end{equation}
\vspace*{-0.1cm}
\vspace*{-0.65cm}
\section{$K^0-\overline{K}^0$}
\begin{figure}
  \centering
  \subfloat[]{\label{fig:Bi-plateau}\includegraphics[scale=0.48]{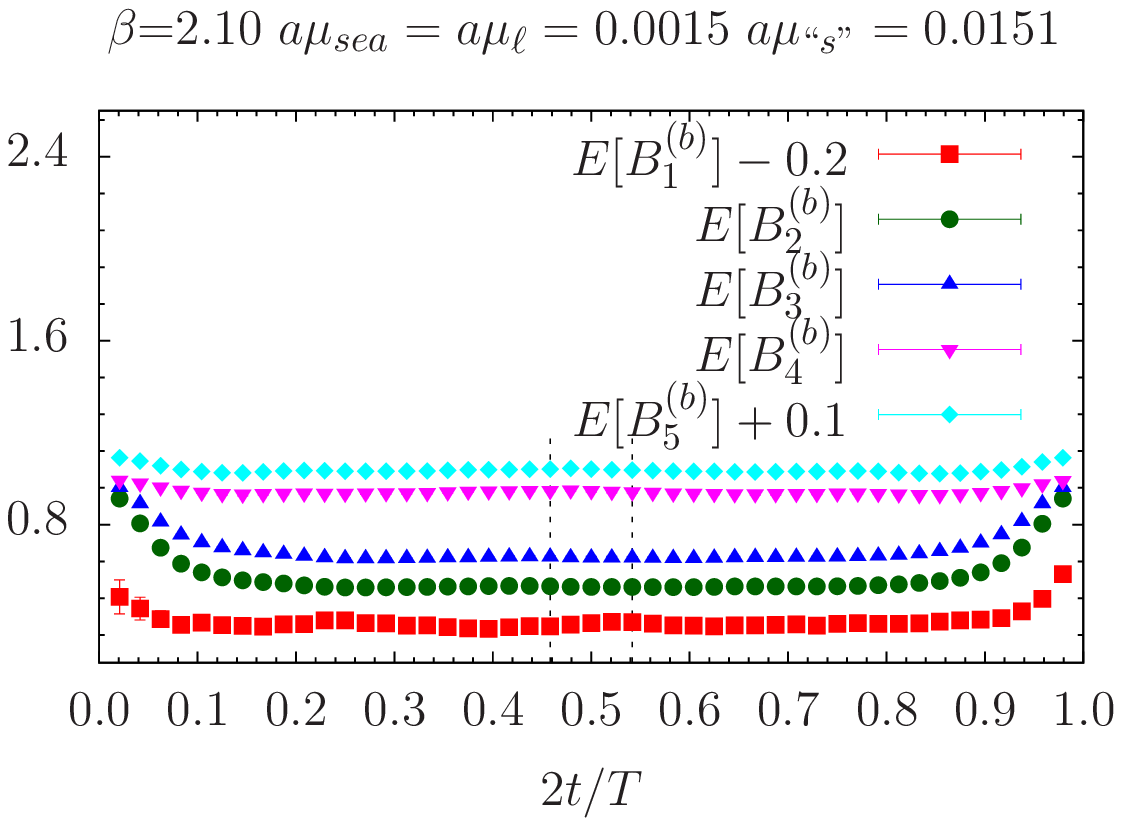}}\hspace{1.5cm}%\hfill               
  \subfloat[]{\label{fig:B1-CL}\includegraphics[scale=0.48]{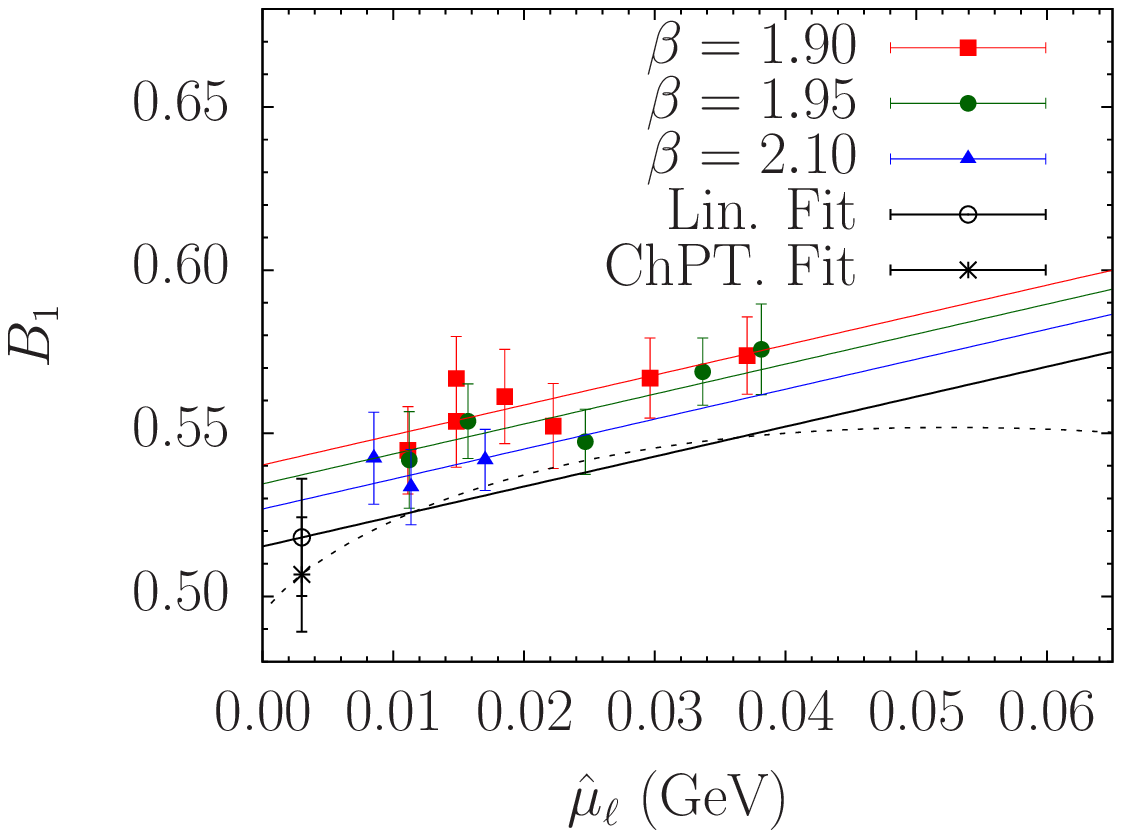}}
  \vspace*{-0.3cm}
 \caption{ (a) $B_i$ plateaux  vs $2t/T$  at $\beta=2.10$ and $(a\mu_{\ell},a\mu_{h})=(0.0015,0.0151)$. The vertical dotted  lines delimit the plateaux region.  (b) Chiral and continuum extrapolation of $B_1$ $K^0$ parameter renormalized in ${\overline{MS}}$ scheme at 3 GeV. $\hat{\mu}_{\ell}$ is the  quark mass renormalised in  ${\overline{MS}}$ at 3 GeV.  The full black line is the continuum limit curve and the dashed black line is the NLO ChPT continuum limit curve.} 
\end{figure}

\vspace*{-0.3cm}
The lattice estimators of bare $B_i$ parameters are obtained from the plateaux of the ratios 
\vspace*{-0.3cm}
\begin{equation}
\begin{array}{ccc}
E[B_{1}^{(b)}](x_{0})=\dfrac{C_{1}(x_{0})}{C_{AP}(x_{0})C'_{AP}(x_{0})}, &  \,\,\,\,\,\,\,\,\,\,\,\,\,\,\,\,\,\,\,\,\,\, & E[B_{i}^{(b)}](x_{0})=\dfrac{C_{i}(x_{0})}{C_{PP}(x_{0})C'_{PP}(x_{0})}\end{array} ,
\end{equation}
\vspace*{-0.1cm}
at times $x_0$ such that $y_0\ll x_0 \ll y_0+T_{\textrm{sep}}$ where $T_{\textrm{sep}}$ is the separation between the two pseudoscalar meson walls. For $K^0-\overline{K}^0$ we fix $T_{\textrm{sep}}=T/2$. The involved correlators are defined as in  \cite{Bertone:2012cu} 
\vspace*{-0.3cm}
\begin{equation}
\begin{array}{lcl}
C_{i}(x_{0})={\displaystyle \sum_{\vec{x}}}\langle\mathcal{P}_{y_{0}+T_{\textrm{sep}}}^{43}Q_{i}(\vec{x},x_{0})\mathcal{P}_{y_{0}}^{21} \rangle , \\
C_{PP}(x_{0})={\displaystyle \sum_{\vec{x}}}\langle P^{12}(\vec{x},x_{0})\mathcal{P}_{y_{0}}^{21}\rangle , &  & C_{AP}(x_{0})={\displaystyle \sum_{\vec{x}}}\langle A^{12}(\vec{x},x_{0})\mathcal{P}_{y_{0}}^{21}\rangle ,\\
C'_{PP}(x_{0})={\displaystyle \sum_{\vec{x}}}\langle\mathcal{P}_{y_{0}+T_{\textrm{sep}}}^{43}P^{34}(\vec{x},x_{0})\rangle , &  & C'_{AP}(x_{0})={\displaystyle \sum_{\vec{x}}}\langle\mathcal{P}_{y_{0}+T_{\textrm{sep}}}^{43}A^{34}(\vec{x},x_{0}) \rangle ,
\end{array}
\end{equation}
\vspace*{-0.1cm}
where $\mathcal{P}$ are the pseudoscalar meson sources
\vspace*{-0.3cm}
\begin{equation}
\begin{array}{ccc}
\mathcal{P}_{y_{0}}^{21}={\displaystyle \sum_{\vec{y}}\bar{q}_{2}(\vec{y},y_{0})\gamma_{5}q_{1}(\vec{y},y_{0})}, &  & \mathcal{P}_{y_{0}}^{43}={\displaystyle \sum_{\vec{y}}\bar{q}_{4}(\vec{y},y_{0}+T_{\mbox{\textrm{sep}}})\gamma_{5}q_{3}(\vec{y},y_{0}+T_{\textrm{sep}})}\end{array} ,
\end{equation}
and 
\vspace*{-0.3cm}
\begin{equation}
P^{ij}=\bar{q}_i\gamma_5q_{j}, \,\,\, A^{ij}=\bar{q}_i\gamma_0\gamma_5q_{j} .
\end{equation}
\vspace*{-0.1cm}
In figure \ref{fig:Bi-plateau} we display the quality of the $B_i$ plateaux  at $\beta=2.10$ and the smallest value of the light quark mass. 
Chiral and continuum extrapolations are carried out in a combined way. As an example, in figure \ref{fig:B1-CL} we show the combined fit for the $B_1$ $K^0$ parameter, renormalized in the $\overline{MS}$ scheme at 3 GeV, against the re\-nor\-ma\-li\-zed  light quark mass.

Alternatively, we can consider the matrix elements ratio
\vspace*{-0.3cm}
\begin{equation}
R_{i}=\dfrac{\langle\overline{K}^{0}|Q_{i}|K^{0}\rangle}{\langle\overline{K}^{0}|Q_{1}|K^{0}\rangle} .
\end{equation}
\vspace*{-0.1cm}
 as first proposed in \cite{Donini:1999nn}. Bare $R_i$ parameters are obtained from the asymptotic time behaviour of
\vspace*{-0.3cm}
\begin{equation}
E[R_{i}^{(b)}](x_{0})=\dfrac{C_{i}(x_{0})}{C_{1}(x_{0})} .
\end{equation}
\vspace*{-0.1cm}
As in \cite{Bertone:2012cu,Babich:2006bh} we choose to evaluate the rescaled renormalized quantity defined as
\vspace*{-0.3cm}
\begin{equation}
\tilde{R}_{i}=\left(\dfrac{f_{K}}{m_{K}}\right)^2_{\textrm{exp}}\left[\dfrac{M^{12}M^{34}}{F^{12}F^{34}}\dfrac{Z_{ij}}{Z_{11}}R_{j}^{(b)}\right]_{\textrm{latt}} ,
\label{eq:Rtilde}
\end{equation}
\vspace*{-0.1cm}
in order to compensate for the chiral vanishing of the $\langle \overline{K}^0| Q_1 |K^0\rangle$ matrix elements and reduce the lattice artefacts due to the different lattice discretizations of the kaon mesons. In Equation (\ref{eq:Rtilde}) we have normalized with the ratio of the experimental quantities $f_K^{\textrm{exp}}=156.1$ MeV and  $m_K^{\textrm{exp}}=494.4$ MeV. Notice that in the continuum limit the quantity $\tilde{R}_i$ of Equation \ref{eq:Rtilde} provides the right estimate for the ratio of the renormalized matrix elements. 

\begin{figure}
\centering
\subfloat[]{\label{fig:Ri-plateau}\includegraphics[scale=0.48]{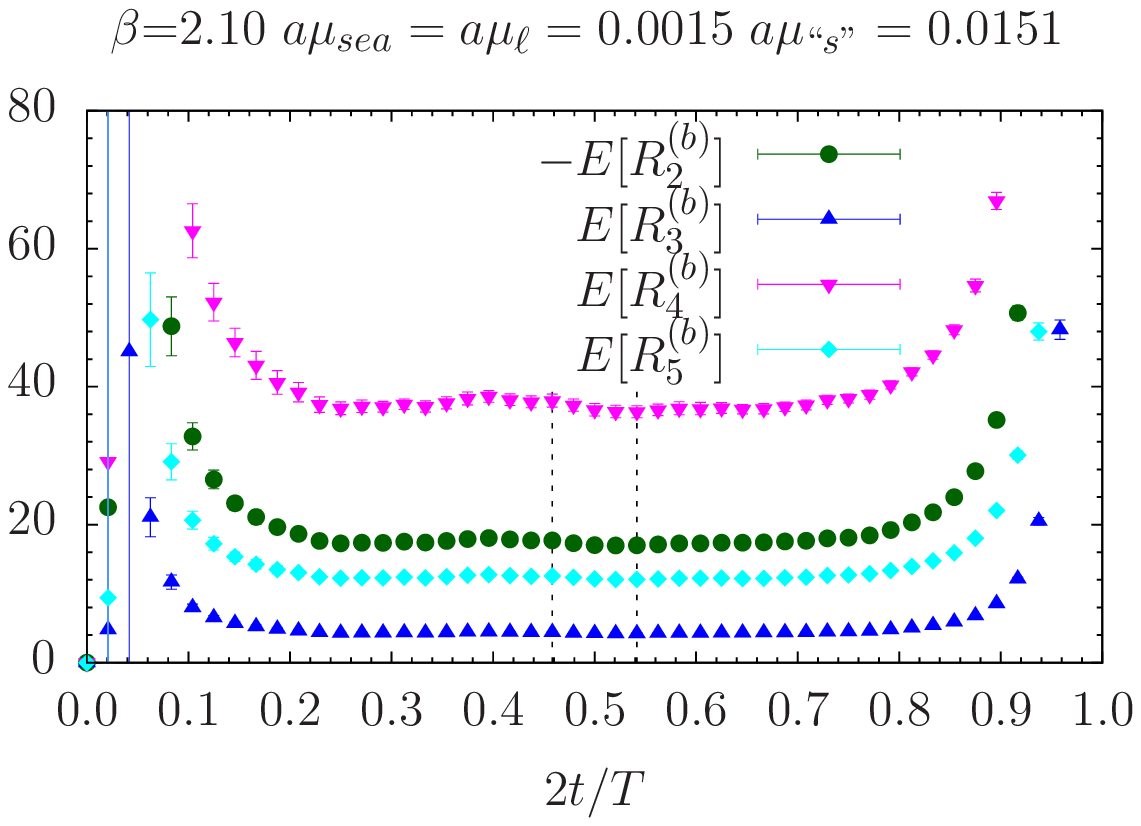}}\hspace{1.5cm}%\hfill               
\subfloat[]{\label{fig:R3-CL}\includegraphics[scale=0.48]{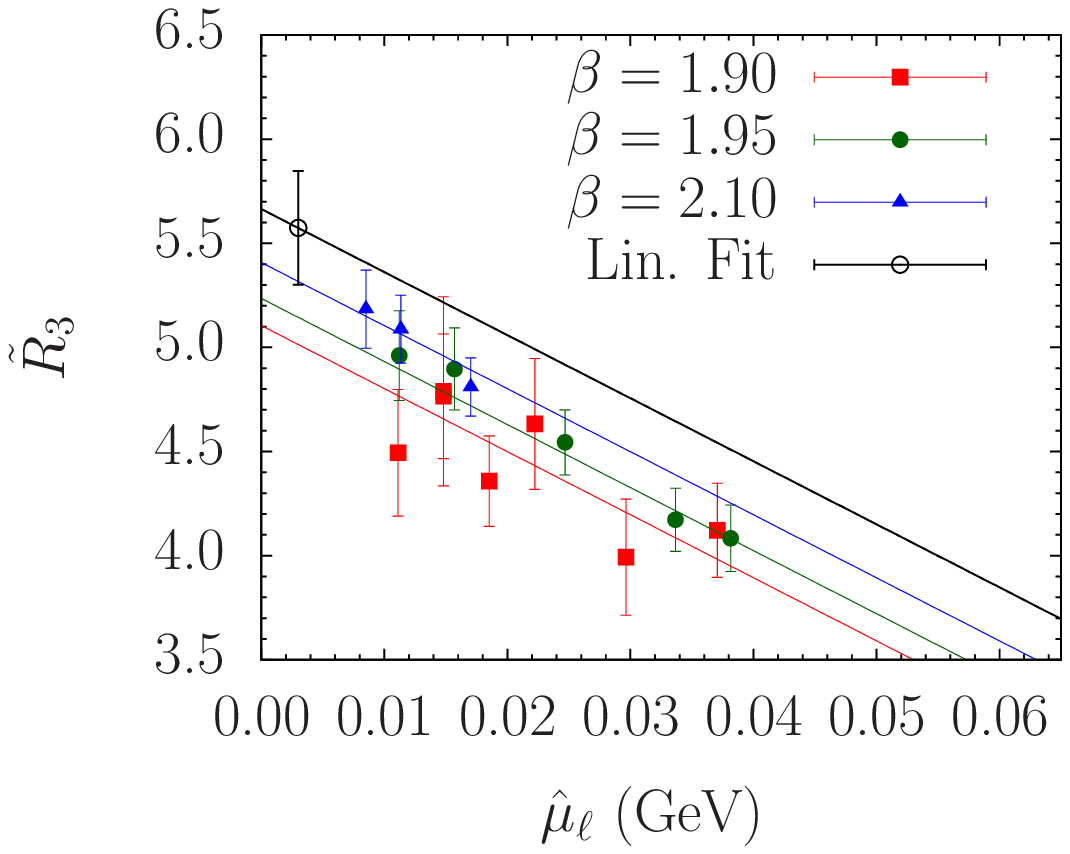}}
\vspace*{-0.3cm}
 \caption{ (a)  $R_i$ plateaux vs $2t/T$  at $\beta=2.10$  lattice and $(a\mu_{\ell},a\mu_{h})=(0.0015,0.0151)$. (b) Chiral and continuum extrapolation of $R_3$ $K^0$ parameter renormalized in ${\overline{MS}}$ scheme at 3 GeV. } 
\end{figure}

Figure \ref{fig:Ri-plateau} is an example of the plateaux quality of the four-fermion operator ratios and figure \ref{fig:R3-CL} shows the combined chiral and continuum fit for the ratio $\tilde{R}_3$ against the renormalized light quark mass. 
In table \ref{tab:BK-results} we gather our final continuum results for $B_i$ and $R_i$ in the $\overline{MS}$ scheme of \cite{Buras:2000if} at 3 GeV with their total error. The systematic error, resulting from discretization effects, the chiral fit and the renormalization procedure, is added in quadrature to the statistical uncertainty. 
\vspace*{-0.3cm}
\vspace*{-0.5cm}
\begin{center}
\begin{table}
\begin{centering} 
\scalebox{0.88}{
\begin{tabular}{|ccccc|}
\hline 
$B_{1}$ & $B_{2}$ & $B_{3}$ & $B_{4}$ & $B_{5}$\tabularnewline
\hline 
0.51(02) & 0.46(02) & 0.81(05) & 0.76(03) & 0.47(04)\tabularnewline
\hline 
 & $R_{2}$ & $R_{3}$ & $R_{4}$ & $R_{5}$\tabularnewline
\hline 
 & -15.1(6) & 5.4(3) & 30.1(1.5) &6.0(3) \tabularnewline
\hline 
\end{tabular}
}
\par\end{centering}{\small \par}
\vspace{-0.15cm}
\caption{\label{tab:BK-results}Continuum limit results for $B_i$ and $R_i$ parameters of the $K^0-\overline{K}^0$ system renormalized in the $\overline{MS}$  scheme  of \cite{Buras:2000if} at 3 GeV.} 
\vspace{-0.25cm}
\end{table} 
\end{center}

\section{$D^0-\overline{D}^0$}
\vspace*{-0.3cm}
$B_i$ and $R_i$ parameters for the $D^ 0-\overline{D}^0$ oscillations can be determined following a similar stra\-te\-gy. However, due to the experimental uncertainty on $f_D$ we modify Equation (\ref{eq:Rtilde}). The re\-nor\-ma\-li\-zed $R_i$ parameters for $D^ 0-\overline{D}^0$ are defined as
\vspace*{-0.3cm}
\begin{equation}
\tilde{R}_{i}=\left(\dfrac{1}{m_{D}^2}\right)_{\textrm{exp}}\left[M^{12}M^{34}\dfrac{Z_{ij}}{Z_{11}}R_{j}^{(b)}\right]_{\textrm{latt}} .
\end{equation}
\vspace*{-0.1cm}

Previous exploratory studies with the ETMC $N_f=2$ data show that using Gaussian smeared sources and choosing a time separation between meson sources smaller than $T/2$ is crucial for quark masses around the physical charm and above \cite{Carrasco:2012dd,Carrasco:2013zta}. In particular, we set $T_{\textrm{sep}}=18$ at $\beta=1.9$, $T_{\textrm{sep}}=20$ at $\beta=1.95$ and $T_{\textrm{sep}}=26$ at $\beta=2.10$.

\begin{figure}
  \centering
  \subfloat[]{\label{fig:BiD-plateau}\includegraphics[scale=0.48]{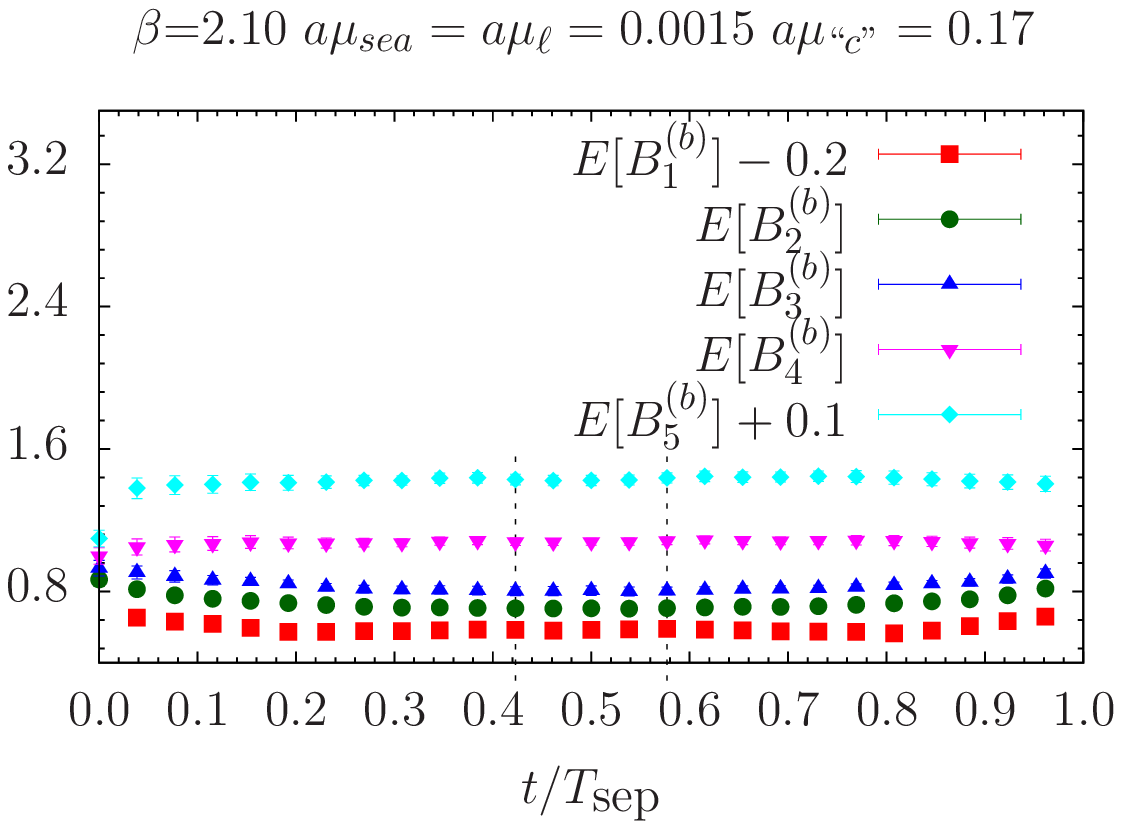}}\hspace{1.5cm}%\hfill               
  \subfloat[]{\label{fig:B2-CL}\includegraphics[scale=0.48]{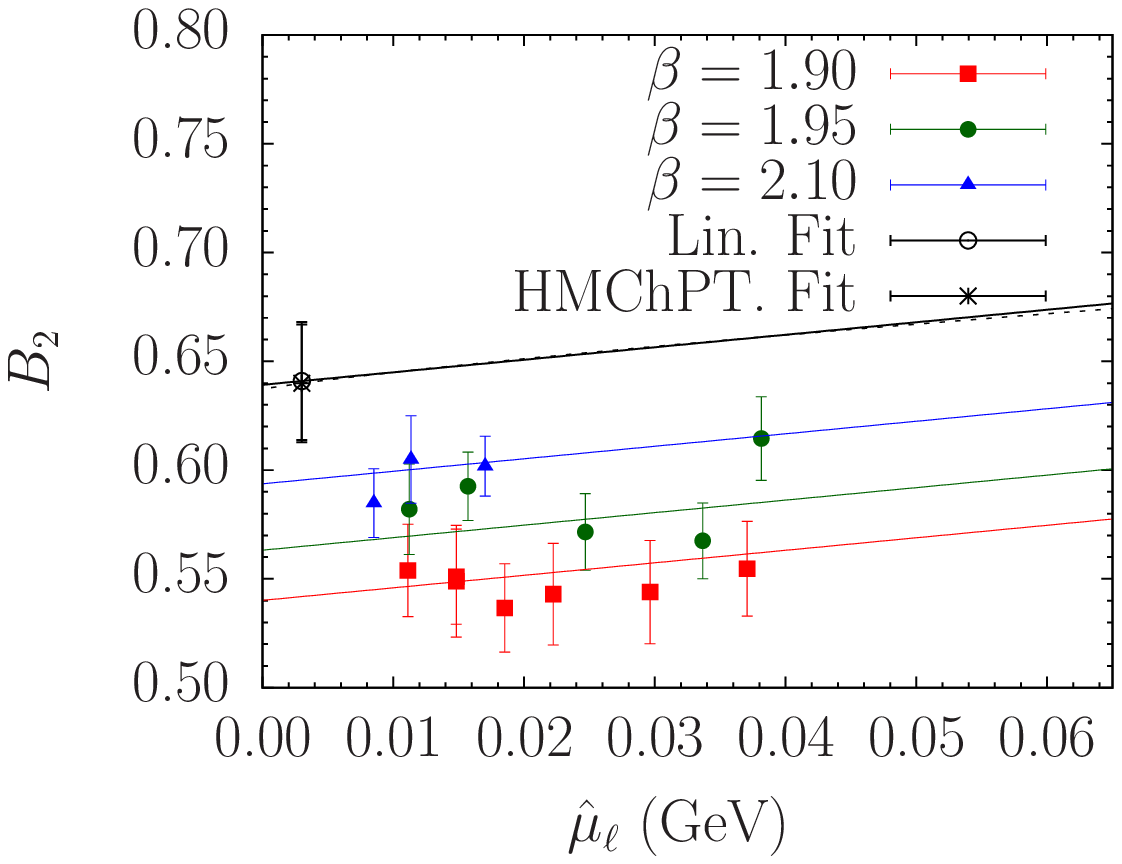}}
  \vspace*{-0.3cm}
 \caption{ (a) $B_i$  plateaux vs $t/T_{\textrm{sep}}$  at $\beta=2.10$ and $(a\mu_{\ell},a\mu_{h})=(0.0015,0.17)$. The vertical dotted lines delimit the plateaux region.  (b) Chiral and continuum extrapolation of $B_2$ $D^0$ parameter renormalized in ${\overline{MS}}$ scheme at 3 GeV. The full black line is the continuum limit curve while the dashed black line represents the continuum limit curve in the case of a NLO HMChPT ansatz.} 
\end{figure}

\begin{figure}
\centering
\subfloat[]{\label{fig:RiD-plateau}\includegraphics[scale=0.48]{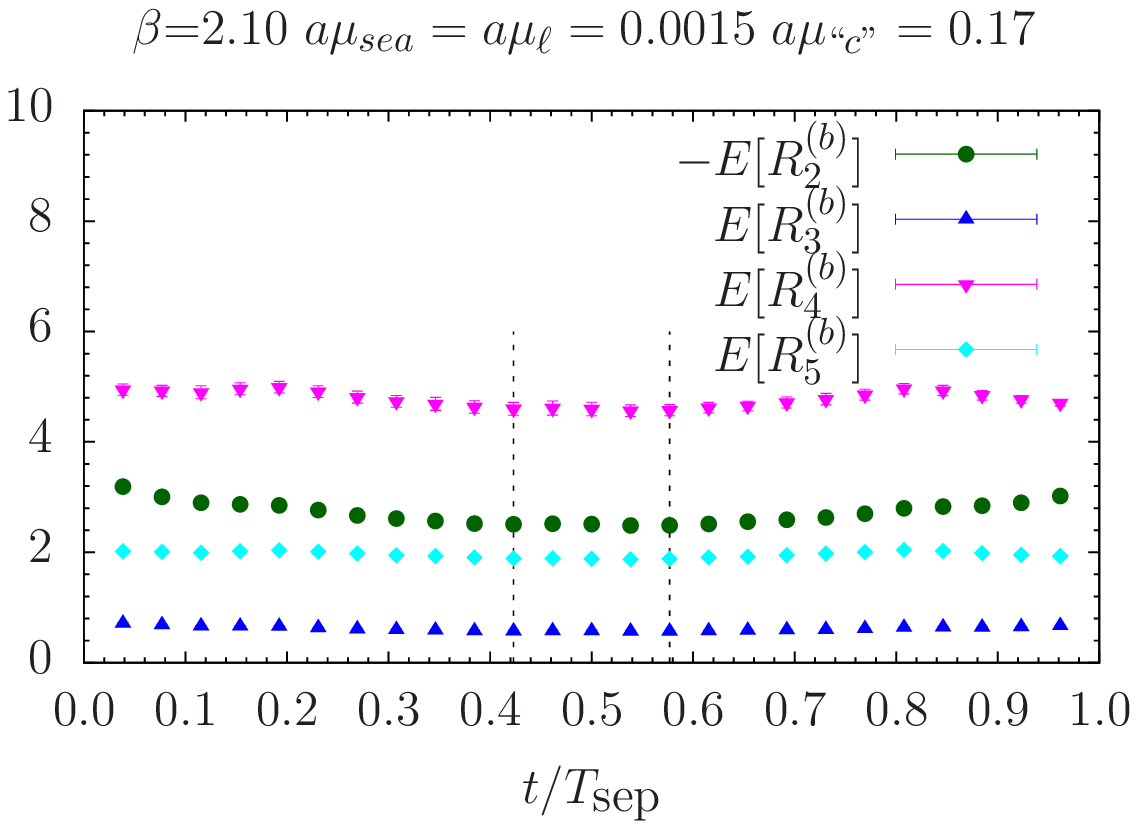}}\hspace{1.5cm}%\hfill               
\subfloat[]{\label{fig:R3D-CL}\includegraphics[scale=0.48]{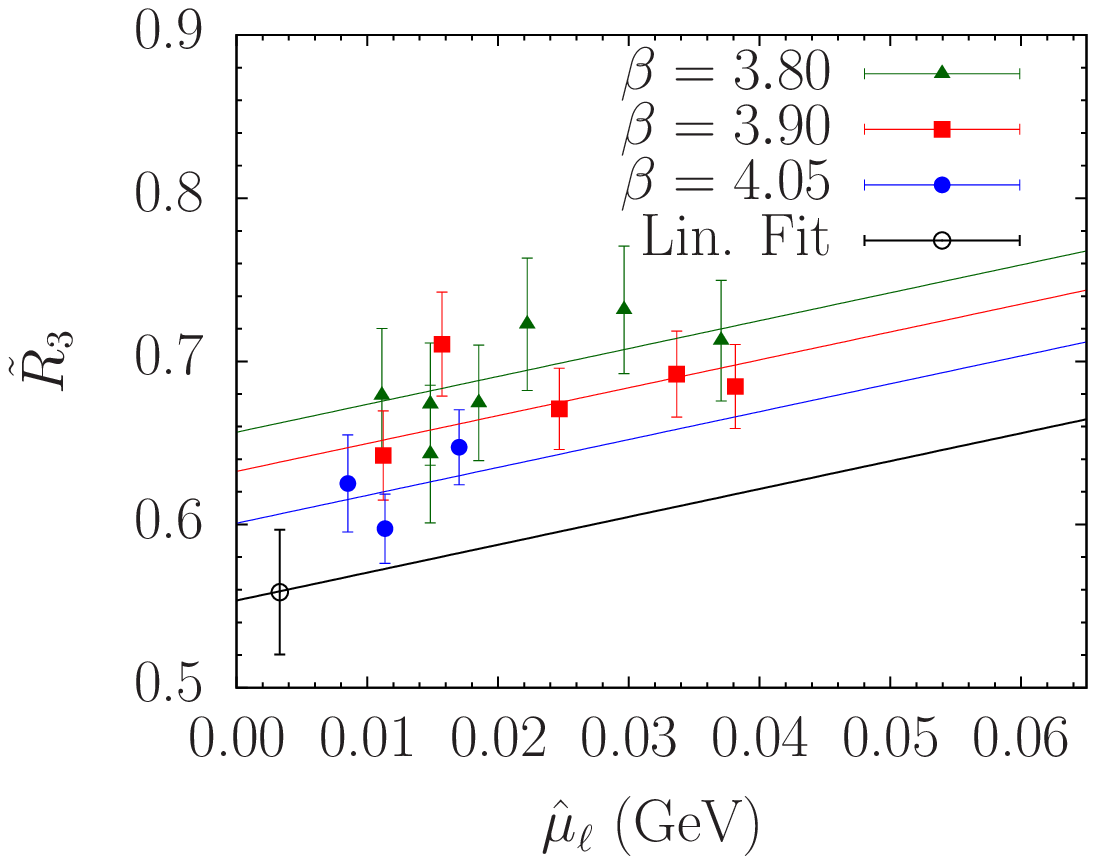}}
\vspace*{-0.3cm}
 \caption{ (a)$R_i$ plateaux vs $t/T_{\textrm{sep}}$ at $\beta=2.10$ lattice and $(a\mu_{\ell},a\mu_{h})=(0.0015,0.17)$. (b) Chiral and continuum extrapolation of $R_3$ $D^0$ parameter renormalized in ${\overline{MS}}$ scheme at 3 GeV. } 
\end{figure}

%This improvement allows us to extract the ground state with more confidence and precision in a wider time interval.  
For illustration, in figure \ref{fig:BiD-plateau} and \ref{fig:RiD-plateau}  we display the plateaux quality for $B_i$ and $R_i$ respectively at $\beta=2.10$   and $(a\mu_{\ell},a\mu_{h})=(0.0015,0.17)$. Figures \ref{fig:B2-CL} and  \ref{fig:R3D-CL} show  examples of the chiral and continuum extrapolation for $B_2$ and $R_3$.     
Finally, in table \ref{tab:BD-results} we collect our final results for $B_i$ and $Ri$ in the $\overline{MS}$ scheme of \cite{Buras:2000if} at 3 GeV. 
    
Using as inputs $R_i$, the value of $B_1$ and the renormalized quark masses one can compute indirectly the $B_i$ ($i=2,3,4,5)$ parameters. The indirect evaluation leads to  results compatible within errors with the results shown in table \ref{tab:BK-results} and table \ref{tab:BD-results} but with larger uncertainties. 
%$B_i$ parameters can be computed in an indirect way once $B_1$ and the renormalized quark masses are known. 
      
%Physical values are obtained by interpolating data in $\mu_{``c"}$  to the physical value $\mu_{c}$  while chiral and continuum extrapolations are carried out simultaneously. 
\vspace*{-0.3cm}
\vspace*{-0.5cm}     
\vspace*{-0.3cm}
\begin{center}
\begin{table}
\begin{centering}
\scalebox{0.88}{
\begin{tabular}{|ccccc|}
\hline 
$B_{1}$ & $B_{2}$ & $B_{3}$ & $B_{4}$ & $B_{5}$\tabularnewline
\hline 
 0.76(04)&0.64(02) &1.02(07)  &0.92(03)  &0.95(05)\tabularnewline
\hline 
 & $R_{2}$ & $R_{3}$ & $R_{4}$ & $R_{5}$\tabularnewline
\hline 
& -1.67(09)& 0.53(05)&3.00(15) &1.02(07) \tabularnewline
\hline 
\end{tabular}
}
\par\end{centering}{\small \par}
\vspace*{-0.3cm}
\caption{\label{tab:BD-results}Continuum limit results for $B_i$ and $R_i$ parameters of the $D^0-\overline{D}^0$ system renormalized in the $\overline{MS}$  scheme  of \cite{Buras:2000if} at 3 GeV.} 
\vspace*{-0.3cm}
\end{table} 
\end{center}

\textbf{Acknowledgements} CPU time was provided by the PRACE Research Infrastructure resource JUGENE based in Germany at Forschungzentrum Juelich (FZJ) under the projects PRA027  "QCD Simulations for Flavor Physics in the Standard Model and Beyond" and PRA067 "First Lattice QCD study of B-physics with four flavors of dinamical quarks" and by the Italian SuperComputing Resource Allocation (ISCRA) under the class A project HP10A7IBG7 "A New Approach to B-Physics on Current Lattices" and the class C project HP10CJTSNF "Lattice QCD Study of B-Physics" at the CINECA supercomputing service.

\end{document}